%% file: main.tex
\documentclass[a4paper,11pt]{article}
\pdfoutput=1 

\usepackage{jcappub} 

\usepackage[T1]{fontenc} 

\usepackage{orcidlink}

\title{\boldmath Using Active Learning to Improve Quasar Identification for the DESI Spectra Processing Pipeline}

\include{qnet_authors}

\emailAdd{dylanag@uci.edu}

\abstract{The Dark Energy Spectroscopic Instrument (DESI) survey uses an automatic spectral classification 
pipeline to classify spectra. QuasarNET is a convolutional
neural network used as part of this pipeline originally trained using data from the
Baryon Oscillation Spectroscopic Survey (BOSS). 
In this paper we implement an active learning algorithm to optimally select spectra to use
for training a new version of the QuasarNET weights file using only DESI data, with the goal of improving classification accuracy. 
This active learning algorithm includes
a novel outlier rejection step using a Self-Organizing Map to ensure we label spectra representative
of the larger quasar sample observed in DESI. 
We perform two iterations of the active learning pipeline,
assembling a final dataset of 5600 labeled spectra, 
a small subset of the approximately 1.3 million quasar targets in DESI's Data Release 1. 
When splitting the spectra into training and validation 
subsets we achieve similar performance to the previously trained weights file in completeness and purity calculated on the 
validation dataset but do so with less than one tenth of the amount of training data. 
The new weights also more consistently classify objects in the same way when used on unlabeled data
compared to the old weights file. In the process of improving QuasarNET's classification accuracy
we discovered a systemic error in QuasarNET's redshift estimation and used our findings to improve
our understanding of QuasarNET's redshifts.}

\begin{document}
\maketitle
\flushbottom

\section{Introduction}

The Dark Energy Spectroscopic Instrument (DESI) is an ongoing spectroscopic survey that aims to 
observe over 40 million objects over the course of five years, with the goal of mapping the expansion
history of the universe \cite{Snowmass2013.Levi, DESI2016a.Science}.
DESI is mounted on the Mayall
4-meter telescope at Kitt Peak National Observatory and will map approximately a third of the night
sky over the course of the survey \cite{DESI2016b.Instr, DESI2022.KP1.Instr}. DESI takes an exposure
of up to 5000 spectra at a time, a significant increase in data collection 
over previous spectroscopic surveys \cite{FocalPlane.Silber.2023, Corrector.Miller.2023, 2024AJ....168..245P}. To date \cite{2024arXiv241112020D}
approximately one million spectra were released publicly in the DESI Early Data Release (EDR) \cite{DESI2023a.KP1.SV, DESI2023b.KP1.EDR}, 
followed by more than 18 million spectra in Data Release 1 (DR1). 

The first data release of the DESI instrument \cite{DESI_collab_I} includes key science papers presenting
the cosmological results of the first year of data collection. These results span a variety of
different cosmology probes, and include two point clustering \cite{2024arXiv241112020D}, Baryon Acoustic
Oscillation (BAO) and full shape results from both quasars and galaxies \cite{2024arXiv240403000D, 2024arXiv241112021D}
and BAO results from the Lyman-alpha forest \cite{2024arXiv240403001D}. DESI also
reports cosmological constraints from the BAO results \cite{2024arXiv240403002D} 
and the full shape analysis \cite{2024arXiv241112022D}.

As part of its data collection, DESI employs a state-of-the-art data processing pipeline that automatically classifies each spectrum  \cite{Spectro.Pipeline.Guy.2023, 2023AJ....166..259S},
redrock \cite{redrock},
which uses Principal Component Analysis
templates fit at different trial redshifts. While overall very accurate, redrock is slightly less
accurate when identifying quasars compared to other galaxy sub types. To this end, QuasarNET
was added to the DESI classification pipeline to recover quasars missed by redrock 
\cite{chaussidon_target_2023}. QuasarNET is a convolutional neural network that 
can automatically identify quasars and provide an estimated redshift based on observed emission
line positions \cite{busca_quasarnet_2018, farr_optimal_2020}. 
We refer readers to \cite{chaussidon_target_2023} for a more detailed description of how QuasarNET 
contributes to DESI's classification pipeline.

QuasarNET originally consisted of four convolution layers followed by a fully connected layer. This
fully connected layer was then connected to a series of fully connected layers of 26 outputs each. 
The number of fully connected layers is determined by how many quasar emission lines QuasarNET is trained on. Each output fully connected layer represents one line-finder n=unit. 
The first 13 outputs of each line-finder unit is a confidence value that
the given emission line is present within that portion or ``box'' subset of the wavelength range, 
a coarse-grained estimate of the line position.
The second 13 outputs are fine-grained estimates of line positions within each of the 13 boxes and
are floating point values between -0.1 and 1.1, where 0 corresponds with the lower edge of the box
and 1 corresponds with the upper edge. Classification of spectra using these outputs is done
by selecting a confidence threshold and a number threshold and assigning a quasar classification
to any spectra where more lines than the number threshold have confidence values 
above the confidence threshold. For example, in DESI any spectrum that has at least one emission
line (number threshold of one) of confidence greater than 0.95 (confidence threshold of 0.95) in QuasarNET's output will be classified as a quasar. 
Classification therefore only only uses the first 13 outputs of each line-finder unit, namely those
that are associated with line confidence.
QuasarNET also provides a redshift estimate for each spectrum, producing an independent redshift 
estimate for each emission line detected in the spectrum. The final reported redshift by QuasarNET is 
then the redshift of QuasarNET’s most confident line. Redshift estimation is done by using the
combined estimated line position (coarse + fine) to estimate a central wavelength of an emission
line, which is then converted to a redshift using the rest-frame wavelength associated with that
line-finder unit. \cite{busca_quasarnet_2018} provides a more in-depth explanation of how
QuasarNET works.

Within the DESI pipeline, QuasarNET is run on \emph{all} spectra, but its results are only
used when redrock classifies a spectrum as a galaxy and QuasarNET classifies the same spectrum as a
quasar with a confidence threshold of 0.95 and a number threshold of 1. When this occurs,
QuasarNET will trigger a rerun of redrock using only the quasar templates
and a uniform flat prior on redshift centered at the QuasarNET redshift with a width of $\Delta z = 0.1$.

Since QuasarNET is a deep neural network, its performance heavily depends on the weights file.
The weights file used in DESI EDR and DR1 analyses was originally trained on data from eBOSS with truth labels
provided by visual inspection (VI) in DR12 \cite{paris_sloan_2017}. The eBOSS 
dataset, however, is a suboptimal training sample for the DESI spectral sample set
for a few core reasons:
\begin{itemize}
    \item BOSS and later eBOSS used a log-linear wavelength grid for spectral extraction, and correspondingly QuasarNET uses a log-linear grid based on the eBOSS grid. The eBOSS grid used a constant $\Delta \log{\lambda} = 10^{-3}$ spacing while the QuasarNET grid was designed with $\Delta \log{\lambda} = 10^{-2}$. DESI, on the other hand, uses a linear wavelength grid, with a constant linear spacing $\Delta \lambda = 0.8$. 
    \item The flux calibration, and by extension the signal to noise ratio, between eBOSS and DESI differ over the course of the wavelength region, which might affect the success rate of QuasarNET on DESI spectra. 
    \item The BOSS spectrograph had only a single dichroic \cite{smee_multi-object_2013}, which means that each spectrum was captured in two parts and then coadded into a single wavelength region. DESI, in comparison, has two dichroics and therefore three different wavelength regions. Due to this design, DESI uses a different sensor in the lowest wavelength region than in the upper two, both of which differ from the BOSS sensors. The noise response of these three sensors differ, but QuasarNET does not include any error or inverse variance information. This means that while correct error handling can account for these differences, QuasarNET might not. 
\end{itemize}

In this paper we modify QuasarNET to better accommodate DESI data, improving its capabilities
and performance. These modifications come in two forms: adjustments to the
network structure, and careful selection of training data, with our novel contribution being the use
of active learning to self select a training sample comprised of only DESI data. Due to the large
amount of candidate data we additionally include a novel outlier rejection step to improve
the efficiency of the dataset selection.

After outlining our modifications to QuasarNET, we outline the active learning methodology and its application to improving
QuasarNET's classification accuracy and purity when coupled with the aforementioned network
changes. Ideally we would like a comparable amount of DESI data to train a new weights file to improve QuasarNET's handling of DESI data. The sheer quantity of spectra observed by DESI, however, 
makes visual inspection of DESI data on a similar scale unfeasible, 
so  by construction any training dataset comprised of DESI data will have significantly fewer spectra 
than that comprised of eBOSS data. 
To combat this reduced amount of training data we apply the active learning methodology to select
an optimal set of data to be labeled that will provide the largest improvement in the network
when trained. We couple this with a novel outlier rejection step designed to further ensure that
the spectra we select to label provide adequate coverage of the possible data space. This methodology
shows significant self improvement, with increase in both completeness and purity after each
active learning step. The final produced DESI trained weights file shows similar performance 
to the previous eBOSS trained weights file, but uses a much smaller training dataset of 
approximately 6000 spectra, only 10\% of the previous training sample. 

On labeled DESI data our new weights file performs similarly to the previously trained weights file,
with a marginal improvement in completeness and similar purity.
We demonstrate this by splitting our selected sample into training and
hold out validation sets. We then demonstrate the new QuasarNET's performance on unlabeled data
by comparing the classification and redshift estimates of repeated exposures of the same objects.
In this test we further demonstrate additional benefits of our new dataset 
with improved repeat classification
accuracy, with QuasarNET more consistently classifying objects the same way in multiple exposures.

Our work in this paper was focused only on classification accuracy, but QuasarNET also provides redshift
estimates that are used in the DESI pipeline. When studying the repeat exposure tests we uncovered
a significant ``oscillation'' problem with QuasarNET redshift estimates. We will outline the issue and its causes, which are not
a result of the active learning methodology presented in this paper. At the end of the paper 
will very briefly describe a solution to the problem that was driven by the understanding
gained in this work.

\begin{figure}[tbp]
\centering 
\includegraphics[width=\textwidth]{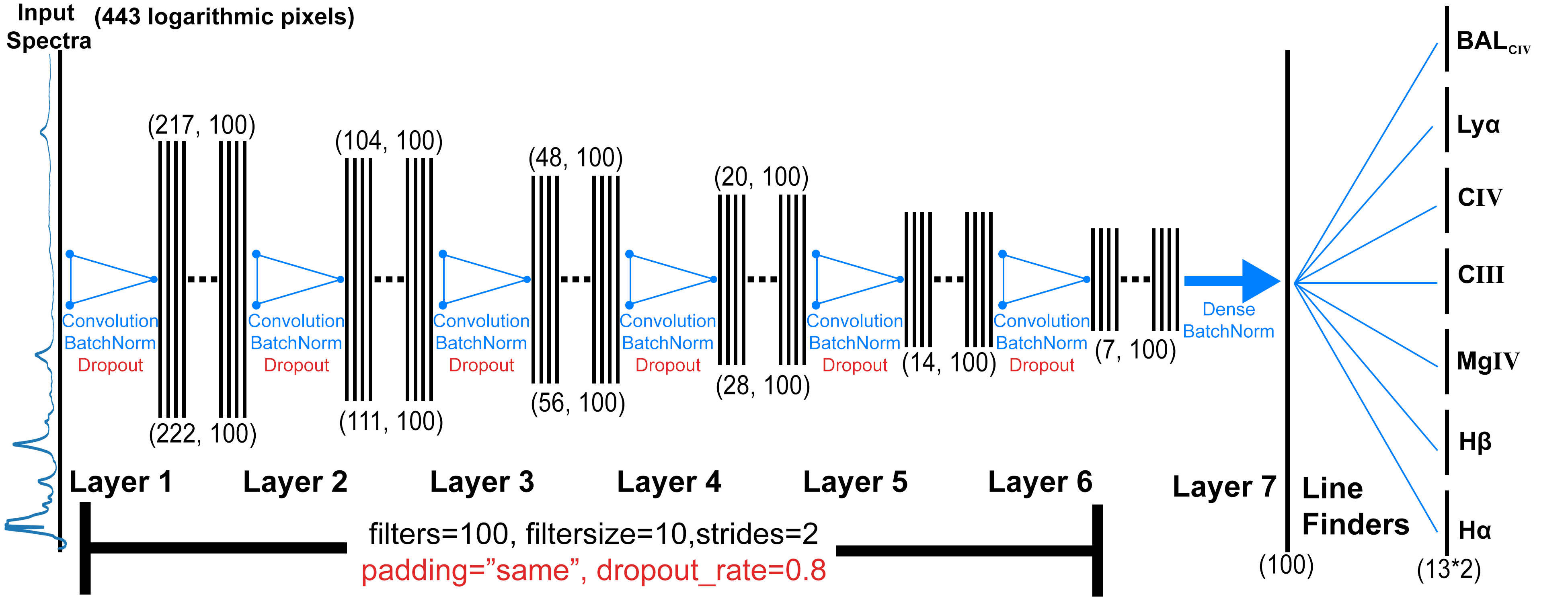}
\caption{\label{fig:new_qnet} QuasarNET architecture, both before and after this work. Along the top are the sizes of the outputs of each layer prior to this work, and below are the numbers after this work. In red along the bottom and at each layer step denote hyperparameters that have been changed by this work. The addition of padding in each convolutional layer allows the inclusion of two additional convolutions, the former structure connected the output of layer 4 directly to the fully connected layer. As before, each line finder is constructed of two fully connected layers, each with 13 outputs, for a total of 26 outputs for each line finder. See text for a more details on the specific changes from prior work.}
\end{figure}

\section{New QuasarNET Structure}

Before selection of a training set we adjust
the QuasarNET input to better accommodate the DESI instrument's output, since QuasarNET was originally designed for eBOSS data. 
We do this by directly modifying the QuasarNET architecture
to better handle the DESI data and its differences from the previous eBOSS data by adding padding,
which increase the amount of convolutions we can include, and by adding dropout, which helps
improves the network's generalizability especially with small training samples.
 

Figure~\ref{fig:new_qnet} shows the QuasarNET model structure prior to this work, originally defined
in \cite{busca_quasarnet_2018}, along the top row. 
The network was constructed of four consecutive convolution layers, each with the same
parameters. The layers have no padding, a stride of two, a convolution kernel width of 10, and a 
filter size of 100. 
The four convolution layers are followed by a fully connected layer of width 100,
which then feeds into 14 parallel individual fully connected layers called line finders, 2 for each line, with 13 outputs each. 
The line finders each produce an independent output estimate of that emission line's position in
the input spectrum, with one block of 13 producing a coarse estimate and the second a fine grained 
estimate on top of the coarse one. 
These layer outputs are then used for classification and redshift estimation,
with an independent result returned for each emission line.
QuasarNET only estimates a redshift if it detects emission lines in the input
spectrum, and although it computes the redshift for all predicted emission lines, in practice
only the redshift estimate from the most confident line is used. 

To improve QuasarNET's performance on DESI data we make two significant changes to the network structure
motivated by the limited amount of DESI training data and differences in data structure and quality.
This new structure is also included in Figure~\ref{fig:new_qnet}, along the
bottom row. Hyperparameters for each layer are printed below each layer, 
with changes highlighted in red across the plot. 
We summarize these changes below.


Firstly, we add dropout layers to each convolution block after the batch normalization step. Dropout
layers randomly select a subset of input units and prevent them from being trained in a given training
epoch, repeating this procedure anew at each epoch. 
Dropout layers have been shown to help 
compensate for a reduced amount of input training data by only training a subset of weights
at each training step, improving the generalization of the network. Additionally, they
have been shown to prevent over fitting, a minor concern when examining loss curves
for the previous eBOSS dataset\cite{hinton_improving_2012, srivastava_dropout_2014}.
Dropout layers have a single hyper parameter corresponding to the frequency at which units are held
out from training, which we set to 0.8, which means that 80\% of the trainable parameters
will be zeroed out and not trained at each epoch. \cite{srivastava_dropout_2014} recommends that
a value of 0.5 tends to be optimal for a wide variety of tasks, but in empirical tests we 
found 0.8 to give the best performance with our specific architecture.

Secondly we add edge padding to the dataset prior to each convolution step. We pad 
each matrix with zeroes using the Tensorflow option ``same'', which pads the matrix with enough
values such that a convolution with a stride of one will output a matrix with the same width
as the input matrix. For our network we pad with zeroes, although in general other values could be used.
We do not change the stride value of the network, leaving it at two. Due to the new padding
each layer will now decrease in width by a factor of two after each convolution, rather than losing half plus the width of the
convolution filter. This
can be seen by comparing the widths of each layer in Figure~\ref{fig:new_qnet} between the old structure
(top) and new structure (bottom). 
This subtle but important difference allows us to add an additional 
two convolution layers before the fully connected layer.

\section{Active Learning and Dataset Selection}
During the first few months of the DESI survey a selection of deep tiles were visually inspected to quantify
the performance of the automatic DESI classification pipeline \cite{alexander_desi_2023, lan_desi_2023}. 
These efforts visually inspected approximately 16000 galaxy targets of various subtypes
and approximately 8500 quasar targets. 
We use a subset of these spectra as an initial training dataset, keeping
all the quasar specific tiles and a small subset of galaxies from the galaxy tiles. 
The small size of this dataset limited the performance of the resultant QuasarNET weights file.

To improve performance we sought to 
label additional DESI spectra to supplement the initial training dataset.
Since the amount of human effort to label and classify spectra is non-trivial, 
we want to minimize the amount of new spectra labeled while maximizing the impact of those spectra 
on improving QuasarNET's performance. We use the first few months of DESI main survey observations, over 3.6 million spectra,
processed in the ``Guadalupe'' data reduction and released as a supplement to DESI's DR1, as a bank
of unlabeled spectra from which we will select spectra to label.
In order to select which of these spectra to label and include in training we use
the active learning algorithm. 
For an in depth and comprehensive exploration of active learning, its implementation, and its
mathematics, we recommend \cite{settles_active_2012}.

Briefly, active learning uses an algorithm's classification uncertainty to quantify which
unlabeled data is most confusing to the network. This confusing data
is then passed to a human expert who labels them for inclusion in a new training run. 
By continuously training on confusing data we can maximize the performance of the algorithm 
with the smallest amount 
of data, as we are continuously filling the network's gaps in knowledge.
When used with a classification algorithm active
learning will tend to pick data points that lie along decision boundaries, i.e. those that
have the highest uncertainty and thus the largest possibility of classification confusion.

Active learning has been used successfully in astronomical research to identify anomalous data to specifically train networks to automatically
identify \emph{interesting} anomalous data \cite{lochner_astronomaly_2021} and to outsource
labeling of galaxy images to citizen scientists to speed classification and improve automatic pipelines \cite{walmsley_galaxy_2020}. These prior applications of active learning in astrophysics 
have focused on image classification networks, which take in a 2-D image and produce a single
classification value. QuasarNET, however, does not directly do classification, nor does it use 2-D images,
instead focusing on identifying emission line positions from 1-D spectra 
with classification and redshift estimation as
post-processing on the raw network outputs. 
We believe that for these reasons our application of active learning is novel in scope.

Since QuasarNET is a line finder, only outputting estimated line positions
for each of the lines in the input set, QuasarNET does not output an uncertainty value 
as part of its results. Classification is done by categorizing objects as 
a quasar if they have at least one of the requisite emission lines with a large enough confidence.
In order to generate uncertainty values for use with active learning we use bootstrapping to create
200 separate weights files using the same input dataset. Each of the weights files are trained using 
the same parameters, the only difference is the data used in training each weights file, which is 
bootstrapped such that each network has the same amount of training data, but different (and possibly
duplicated) subsets of the possible data pool. 

Once we have an ensemble of networks we use the networks to classify each spectrum as either a quasar
or not, using a confidence threshold of 0.95. This is the same confidence threshold used in DESI.
When implementing active learning with only two category classification the algorithm will
pick spectra that lie along the 50\% confusion boundary. It is more interesting and informative
to instead use three or more categories, where spectra that have high confusion between two categories have the possibility of being
strongly identified as \emph{not} belonging to the third. 
To emulate this we split the quasar classification into two separate
categories: HIGH-Z and LOW-Z. We use an estimated value of $z=2.1$ as the division between these
categories, as this is the approximate redshift that the Lyman-$\alpha$ line will become visible enough 
in  the DESI spectrograph for QuasarNET to reliably detect. 
An added benefit of this split is that we can incorporate some estimated redshift
information into the active learning algorithm. 
The division boundary between NOT QSO and the two QSO categories 
will be dominated by whether or not the network has detected any emission lines, while the boundary
between the LOW-Z and HIGH-Z QSO categories is dominated by redshift estimation. Once
we have the output of the 200 networks we convert them to confidence values such that each 
potential unlabeled spectrum has three confidence values defined by
\begin{equation}
    C_{\text{CLASS}} = \frac{N_{\text{CLASS}}}{N_{\text{ENSEMBLE}}},
\end{equation}
where $N_{\text{CLASS}}$ is the number of networks that classified the spectrum with the given 
classification and $N_{\text{ENSEMBLE}} = 200$ is the total number of networks in the ensemble. We use these
confidence values to calculate the entropy of each spectrum,
\begin{equation} \label{eq:entropy}
    H = \sum_{CLASSES} -C_i\log_2{C_i},
\end{equation}
which we use to determine which spectra we will visually inspect, using entropy as a confusion metric
for data selection. The choice of $\log_2$ is motivated by the choice to use units of bits, 
which are more easily interpretable. Spectra that have higher entropy have more confusion, lying near or on the
boundaries between two or all three of the target classes.

Traditionally, active learning considers all possible unlabeled data (in this case spectra) 
as potential training data in its initial pass.
Since labeling spectra is a time intensive task, and there is a non-trivial amount of anomalous
or otherwise unusual spectra, we have added an additional and novel outlier rejection
step to the active learning process. By doing so we ensure that spectra chosen to be inspected are
useful for training the network and \emph{also} spectra that are expected to be 
representative of the larger DESI spectral sample, as opposed to outlier or unusual spectra. 
This outlier  rejection step uses a self-organizing map (SOM) to identify similar spectra to each 
other in a two dimensional reduced space. 

A SOM is an unsupervised machine learning algorithm that divides a requested latent space into 
a set number of cells. At each iteration of the algorithm data points that share characteristics
are assigned to the same or nearby cells, such that the final product is a set of weights
that classifies input data into an assigned cell. For a more detailed theoretical review
of how SOMs work, we refer the reader to \cite{kohonen_self-organized_1982}. SOMs have been used 
variously in astrophysics, with some examples including 
directly classifying objects and estimating their redshifts \cite{geach_unsupervised_2012} and 
performing photometric redshift calibration \cite{wright_photometric_2020}. 
In our work we do not use
the SOM directly for redshift estimation or classification, using it only as part of our
active learning pipeline to reject anomalous spectra. We use the package \texttt{SOMVIZ} \cite{SOMVIZ2022} to train our SOM.

\begin{figure}[tbp]
\centering 
\includegraphics[width=\textwidth]{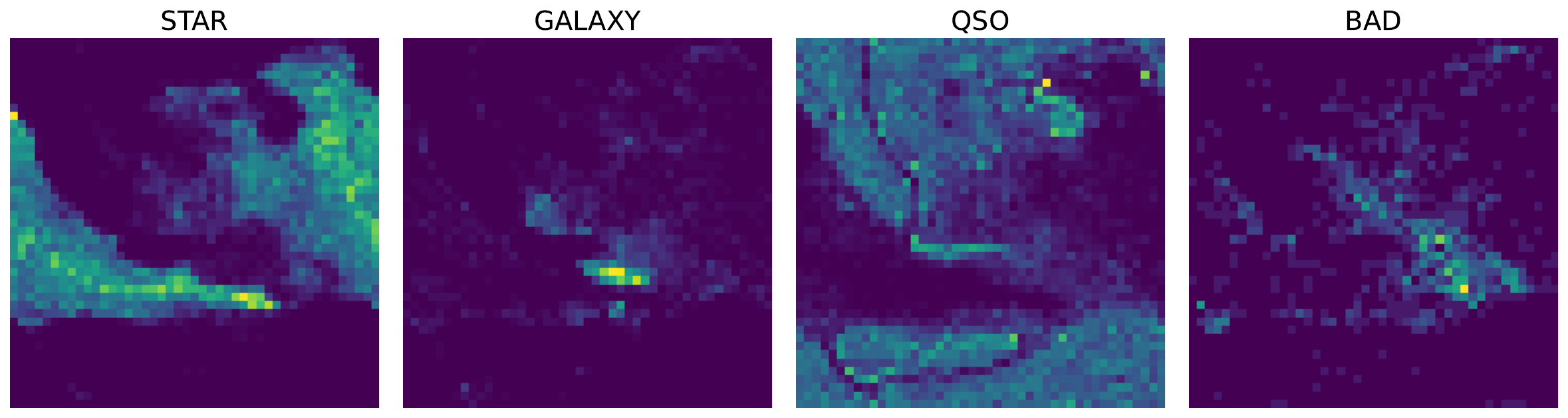}
\caption{\label{fig:som} Two-dimensional representation of the self-organizing map (SOM) used for outlier rejection during active learning. The SOM is square, with 45 cells on each side. In each panel
the color in each cell correlates to how many spectra are assigned to that cell by the SOM. We have
split the SOM into four panels, with each panel representing one of the possible visual inspection
truth labels. By doing this it becomes evident that the SOM has localized similar spectra into self
contained segments of the SOM. Quasars have the largest diversity in our training data, so
take up a larger portion of the SOM.}
\end{figure}

When training our SOM, we use the 
rebinned spectra exactly as they are input to the QuasarNET algorithm, to ensure that it
learns spectral characteristics from the same data as our network. 
The SOM trained and used in our procedure is 
plotted in Figure~\ref{fig:som}. In each panel the color of each cell correlates with how many
spectra are assigned to that cell. We have split the SOM into four panels, with each panel representing
spectra in one of the four possible input truth table labels, to demonstrate how the SOM has associated
similar spectra into localized clumps within the requested two-dimensional space.

When we select a sample of spectra to visually inspect and label, 
we reject any selections whose cell in the SOM has a total number of spectra below a chosen
threshold value. This has the effect of ensuring that the final selected spectra to label 
are localized within the SOM to locations that have a high density of similar spectra. 
For this work, we selected a threshold value equal to the 15th percentile of
spectra counts in all cells.

Once the SOM has removed any potential outlier spectra we select the 1000 objects with the highest
entropy values to be visually inspected and reincorporated into the training dataset. 
We select 1000 spectra to balance 
the impact the new data has on the network against the speed at which we can visually inspect spectra. 
The active learning algorithm we perform can be summarized by the following steps:
\begin{enumerate}
  \item Begin with a set of training data ($T$) and a set of unlabeled potential training data ($U$).
  \item Bootstrap 200 unique subsets of $T$ to train 200 unique versions of QuasarNET.
  \item Run all of $U$ through the bootstrapped QuasarNET versions, generating 200 unique classifications and redshift estimates of all spectra in $U$.
  \item Use a SOM to remove any spectra from $U$ that are outliers, specifically any spectra in bins whose counts are below the 15th percentile of all bins.
  \item De-duplicate spectra of the same objects in $U$, so that the potential dataset consists only of unique objects.
  \item Select 1000 objects from the remainder of $U$ to be visually inspected and labeled.
  \item Label and classify the 1000 selected objects.
  \item Incorporate the newly labeled 1000 objects into $T$ and remove them from $U$.
\end{enumerate}

\noindent After conclusion of step 8 the algorithm can be repeated as many times as desired. We refer to a single iteration of these steps as a ``run''. In this work we perform two runs of active
learning. 

\begin{figure}[tbp]
\centering 
\includegraphics[width=\textwidth]{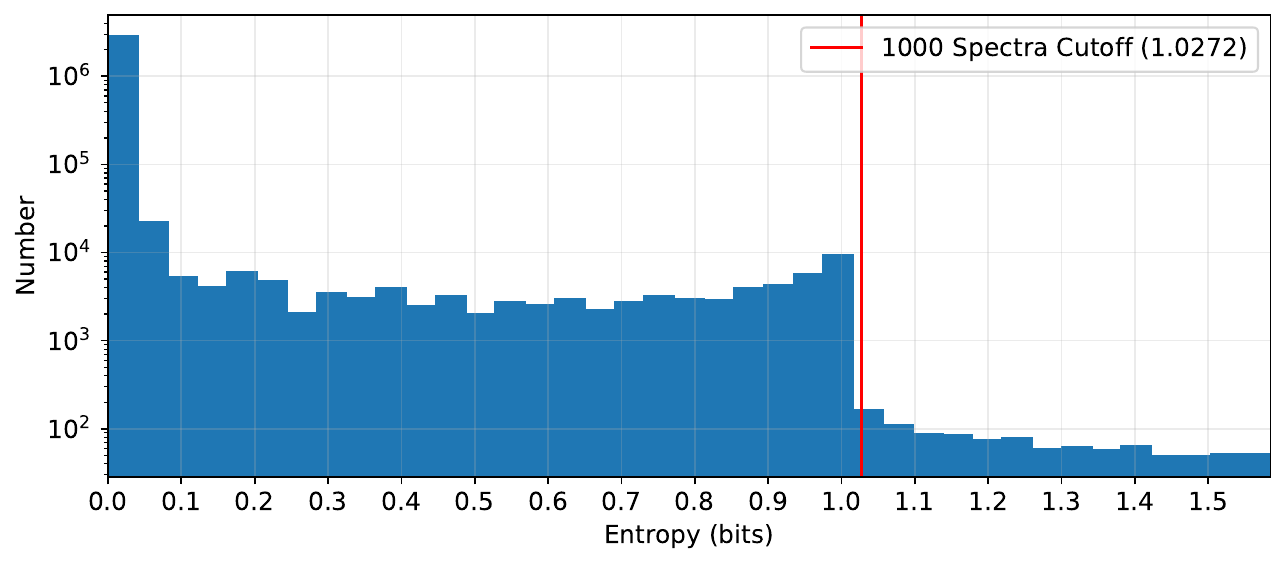}
\caption{\label{fig:entropy} A distribution of entropy values in the unlabeled dataset calculated using the prior weights file. The red line denotes the cutoff above which the 1000 spectra we opted
to VI lie, with the value at which this occurs denoted on the plot. The sharp drop off at around 1 is likely due to the construction of our three classification classes. Since two
of them represent quasar classifications, LOZ and HIZ QSOs, and one does not, it is not equally
likely that spectra will fall into all three classifications based on spectral type alone. }
\end{figure}

The output of a single run of the entire active learning algorithm is represented in Figure 
\ref{fig:entropy}. We start with a potential dataset of 3.6 million spectra 
and apply the SOM based outlier rejection step to eliminate 15.5 thousand spectra 
from the pool of potential training data.
We then apply a cut, removing repeat exposures of the same object to ensure that we have
unique objects to visually inspect rather than multiple exposures of the same objects. 
This reduces the pool of spectra to just below 3 million unique objects. 

In Figure~\ref{fig:entropy} we plot a histogram of the entropy values, defined in Eq. \ref{eq:entropy}, 
of the remaining spectra. A red line indicates the cutoff we selected, above which there are 1000 spectra
that we request to be visually inspected and labeled. 
The plot shows two notable features: first, there is a sharp peak at zero, representing a large set of spectra for which there
is high degree of confidence and agreement in the ensemble classification. 
Second, there is a sharp dropoff near 1 corresponding to the point at which a spectrum
lies at the 50\% confusion boundary between two classes, with no classifications in the third.
Above this point in the histogram are 
spectra that show significant confusion between all three classes, rather than being either
confused between only two classes or highly confident in one and weakly confident in the others.


\section{Results}

We ran the active learning pipeline twice, selecting 1000 spectra at each step to visually
inspect and label. At the initial step we use 3761 DESI spectra from quasar dense tiles (``Deep-field VI'' from Table 1 of \cite{alexander_desi_2023}) plus an 
additional 839 DESI spectra from non-quasar dense tiles (a subset of all spectra in Table 1 of \cite{lan_desi_2023})  which we include to ensure that the dataset 
is not overly quasar dense. These 4600 DESI spectra were visually inspected and the results
validated in \cite{alexander_desi_2023, lan_desi_2023}. In the following section this dataset of
4600 spectra is named \texttt{DESI VI}.
We then augment this dataset with an additional 15000 spectra from the previous 
eBOSS training dataset. We are motivated
to supplement the DESI-only dataset for the first pass because the DESI only dataset does not fully
cover the possible quasar space. We select only 15000 spectra to avoid overwhelming the DESI data
when training the network; the additional spectra are designed to augment the quasar space
covered by the DESI spectra without superseding any properties of the DESI spectra not present
in the eBOSS data. 15000 eBOSS spectra ensures that roughly a quarter of the total dataset for this
training run are DESI spectra. For our initial active learning run, $T$ thus has a size of $19600$ spectra.

For the second active learning step we use only the originally labeled DESI spectra plus the additional
1000 labeled  spectra from the first active learning step to train the networks, excluding the 
supplemental eBOSS data.
At this point we now consider it representative of the DESI quasar observation space, and believe that the 
active learning algorithm will select spectra to fill any deficiencies. This dataset of 5600 spectra 
(the initial 4600 + 1000 newly labeled spectra) will be referred to as \texttt{DESI AL 1}. In the 
second active learning run we will label an additional 1000 spectra. In the second visual inspection campaign
we additionally investigated 100 high-entropy spectra that were rejected as outliers to verify
that the SOM has identified outlier-like spectra. The combination of the initial 4600 spectra, the 2000
active learning spectra, and the 100 outlier spectra (for a total of 6700 spectra) 
will be referred to as \texttt{DESI AL 2}. We will also refer to a fourth dataset in this section,
\texttt{eBOSS DR12}, comprised of the set of approximately 60,000 (eBOSS) spectra used to train 
the previous version of QuasarNET whose performance we are comparing to.

For the visual inspection steps in the active learning pipeline we require that three separate
experts inspect each spectra, assigning an estimated redshift and spectral classification as well
as a ``quality'' rating
that quantifies how confident the inspector is in their
labels. Quality ratings are assigned from zero to four, with four the most confident, and while
each expert makes their own determination of their own confidence they are broadly instructed that
quality zero is reserved for if there is no data. Qualities one and two are generally for uncertain
or insecure redshifts, and three and four are used for secure redshifts.
Each spectra is then inspected by a ``merger'' who resolves discrepancies in the three 
expert's labels and assigns a final merged quality rating. A more detailed description of this procedure is available in \cite{alexander_desi_2023} and \cite{lan_desi_2023}.
QuasarNET only uses spectra whose final merged quality is above 2.5. 

We present the results of this two-step
procedure in three subsections. The first subsection 
includes a short discussion on the inspection of the rejected 100 spectra we visually inspected 
during the second active learning run, justifying their inclusion in the \texttt{DESI AL 2} dataset.
The second and third subsections outline our results: in the second we report results using the full truth dataset 
which we split into training and validation subsets. 
In the third and final subsection we test the final DESI-specific weights file, trained on the
entirety of the truth table rather than a subset. The weights testing used the internal
data release ``Guadalupe'', which is included in DESI's Data Release 1 (DR1),
where we have no truth data but for which we do have a set of metrics that quantify
this weights file's improvement over the previous version. These metrics 
include consistency in classification and redshift uncertainty. In the third section we will also
outline the new understanding of QuasarNET's redshift estimation problems, 
which became apparent when using this newly trained weights file. 

\subsection{Investigation of High Confusion SOM-Rejected Outliers}

\begin{figure}[tbp]
\centering 
\includegraphics[width=\textwidth]{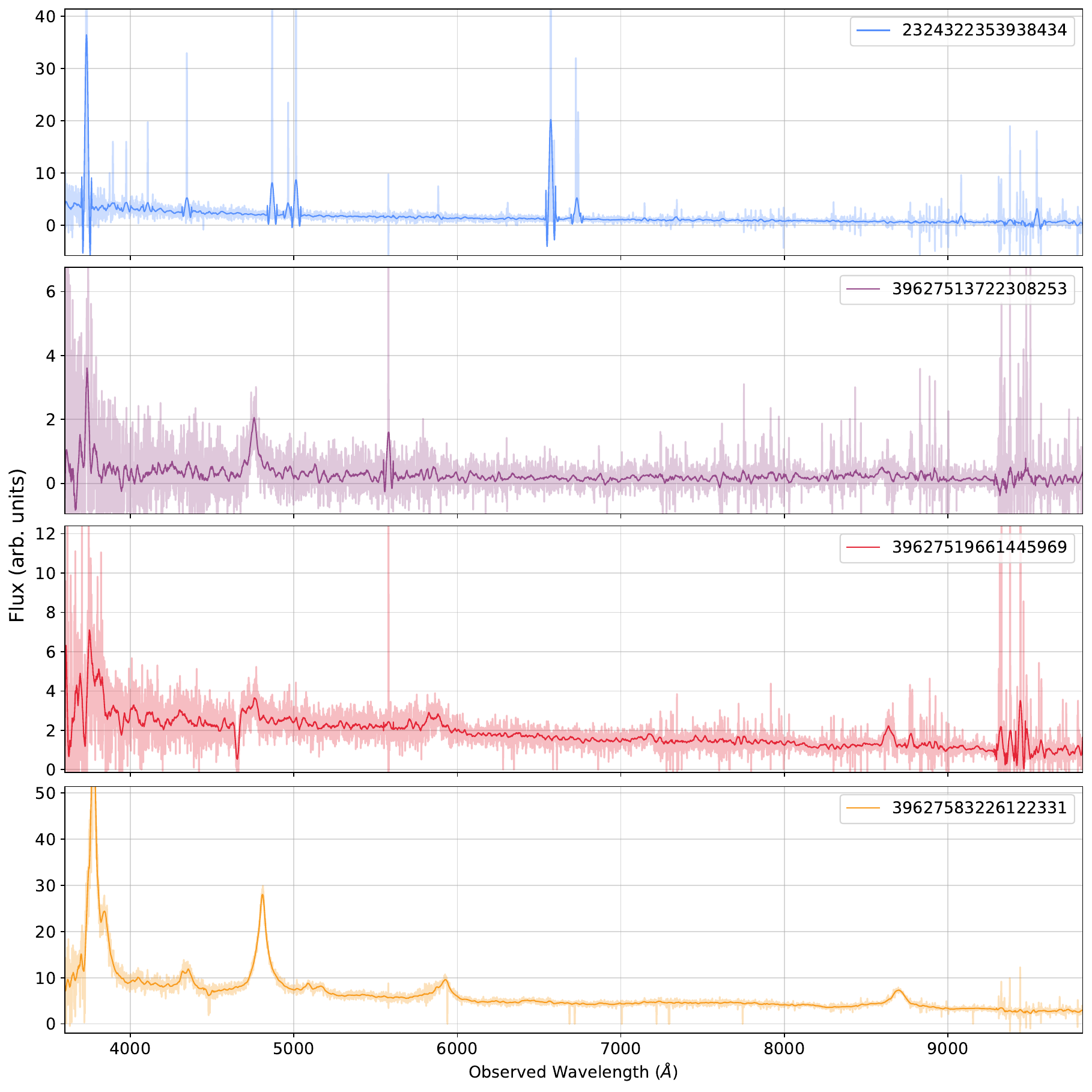}
\caption{\label{fig:example_outliers} An example of four representative high confusion spectra rejected by the SOM as outliers. Each plot shows the raw spectrum from the ``Guadalupe'' reduction as the lighter curve. The darker overplotted line in each plot is the spectrum smoothed with a Savgol filter with window size 75 pixels and polynomial order 3. The top (blue) spectrum is a high SNR and low redshift ($z \lesssim 0.4$ galaxy). The next (magenta) spectrum is a quasar with weak or no CIII emission which is expected around 6000 \AA. The next (red) spectrum is a quasar with narrow CIV emission around 4800 \AA. The bottom (orange) spectrum is an unusually high SNR quasar. The top three spectra all show an unusual high flux pixel near 5600 \AA due to the sky subtraction and calibration.}
\end{figure} 

In this section we will briefly review the visual inspection of the 100 outlier spectra included in the
second visual inspection campaign. These 100 spectra were rejected by the SOM and were investigated to
determine if they were correctly rejected as outliers. These 100 spectra were the highest confusion spectra
in the second active learning run that were rejected by the SOM, and were visually inspected using the
same procedure as the rest of the (non outlier) data.

The SOM rejects spectra if there's too few other similar
spectra in the dataset, so without visual inspection we cannot determine exactly why a spectrum was
rejected. Most if not all of the 100 spectra visually inspected are correctly identified as outliers. 
Broadly many of these outliers are outliers fall into two categories: 
data quality outliers and object type outliers. A few representative spectra are shown in 
Figure~\ref{fig:example_outliers}. These four representative spectra exhibit characteristics
present in all of the 100 inspected outliers. The top three panels (blue, purple and red) 
show object type outliers:
a low redshift galaxy with strong emission lines, a quasar with weak to no CIII emission, and a 
quasar with narrow CIV absorption respectively. The last spectrum is of note because no \textit{broad}
CIV absorption was found in the sample of rejected spectra, only those with narrow absorption. 
The bottom (orange) panel shows a data quality outlier: the quasar has an unusually high SNR
and strong SIV. Even among outliers this is comparatively rare, most of the data quality outliers
are spectra that have such low SNR that identifying spectral features is impossible, if they exist at
all. 

From this sample we concluded that the SOM is accurately rejecting outliers, even if some of the outliers
may be surprising. All four of these representative spectra also represent areas of high confusion for 
QuasarNET: the first three are unusual objects with rare or unusual spectral features, the last
is so high SNR that QuasarNET disagrees on line assignment (leading to redshift confusion, this spectrum
lies along the HIGHZ/LOWZ confusion boundary). Since we were able to generate secure, high quality visually
inspected redshifts for these 100 spectra, we opted to fold them into the \texttt{DESI AL 2} dataset.

\subsection{Results on Labeled Data}

\begin{figure}[tbp]
\centering 
\includegraphics[width=\textwidth]{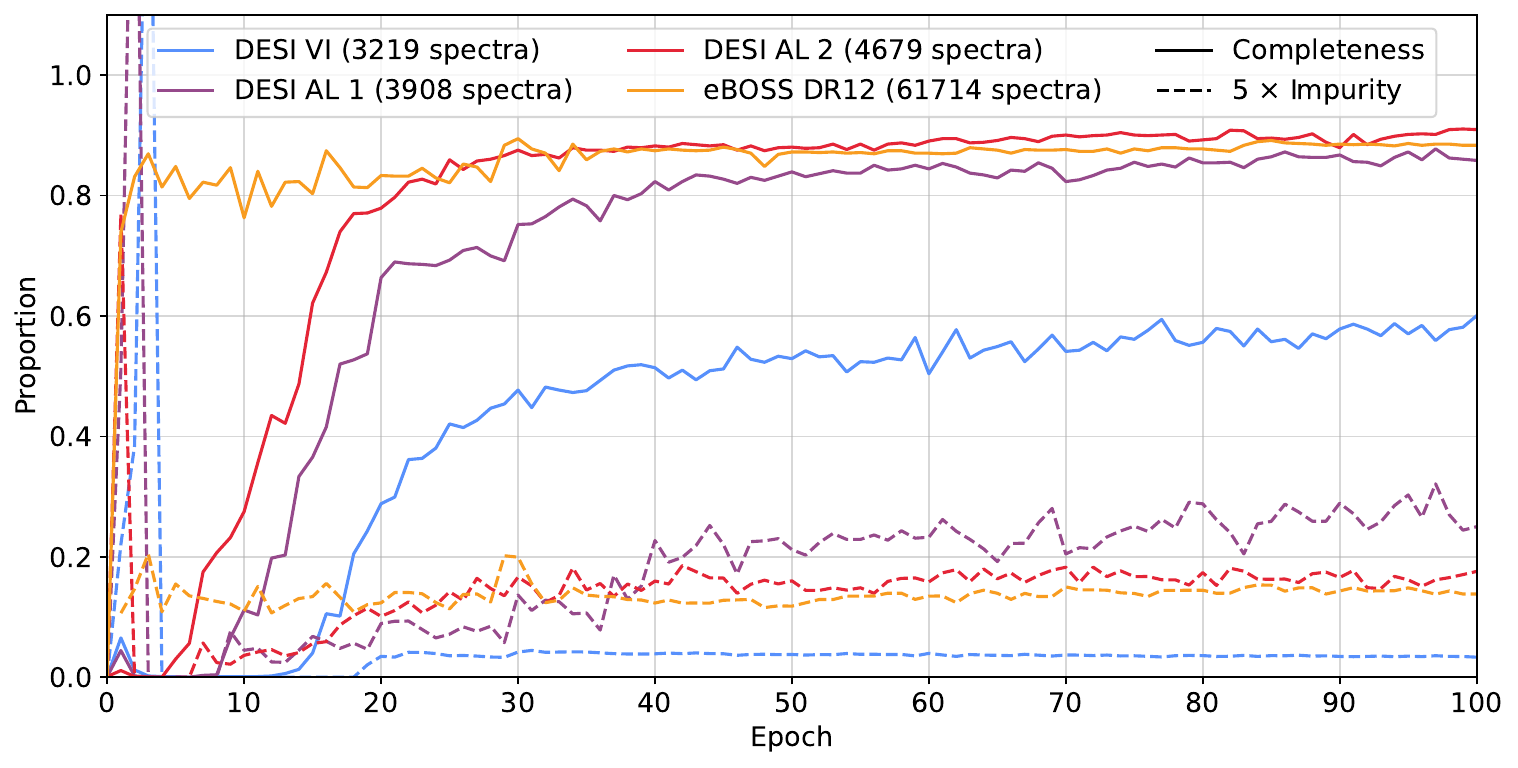}
\caption{\label{fig:purcom}
 Purity and completeness on the validation dataset per epoch of training. Purity is the proportion of classified quasars that are true quasars and completeness is the proportion of true quasars that are correctly classified as quasars. Completeness is plotted in solid curves while in dotted we plot 5~$\times$~impurity for visual clarity. In yellow is a training run using the previous eBOSS based dataset for training; while red, blue and magenta are DESI only datasets. Blue is training on just spectra from DESI Visual Inspection, magenta is after including the results of the first active learning iteration, while red is after the second active learning iteration. When including the additional data determined through active learning we are able to achieve comparable or better performance than the eBOSS trained dataset with less than one tenth of the amount of data. }
\end{figure}

In this section we train a weights file using a subset of the full truth table (\texttt{DESI AL 2}), which includes
the original set of visually inspected quasar tiles (3761 spectra), the additional subset of other galaxy
dense visually inspected tiles (839 spectra), 
and the results of two active learning runs (2100 spectra). The total
number of spectra in the truth table is 6700, although after training we discovered 2 spectra whose data 
processing was corrupted that were affecting training and removed them 
for a final total of 6698 labeled spectra. These two spectra were both visually inspected with
high confidence based on the uncorrupted data, but the presence of data processing corruption
on approximately 1/3rd of the wavelength grid in these two spectra 
made them unsuitable for use in QuasarNET. We split this 
dataset into two subsets such that 70\% of the data is used for training and the other 30\% is used
for testing and validation. The validation dataset is comprised of 2010 spectra, the remaining 6790
are available for use as training data.

In order to validate that the inclusion of the active learning spectra
has improved the network's performance we compare this training to three other training
datasets: \texttt{DESI VI}, \texttt{DESI AL 1}, and \texttt{eBOSS DR12}. The results of these four runs are shown in Figure~\ref{fig:purcom}. \texttt{DESI VI} is shown in blue, \texttt{DESI AL 1} is in magenta,
\texttt{DESI AL 2} is in red and \texttt{eBOSS DR12} is in orange.
For each of the other training schemes we use the same validation dataset (the subset of 
\texttt{DESI AL 2} reserved for validation) excluding from the given training dataset
any spectra that appear in the validation data. The numbers in the plot represent the number of
spectra used to train that network after removing spectra in the validation dataset and after making
the QuasarNET visual inspection quality cuts. For this plot, we use a confidence
threshold of 0.95, declaring that QuasarNET identifies a quasar if it has at least one line with 
a confidence of more than 0.95. 

Completeness, defined as
\begin{equation}
    C = \frac{N_{\text{QSO,PRED}}}{N_{\text{QSO,TRUE}}},
\end{equation}
is plotted in Figure~\ref{fig:purcom} as a solid line per epoch of training and is the proportion of true quasars that are 
correctly identified as such by QuasarNET with 1 meaning that every quasar is correctly identified
by QuasarNET. Purity is the proportion of spectra that are classified as quasars by QuasarNET that are true quasars,
with 1 representing that all predicted quasars are truly quasars. For visual clarity we plot \textit{impurity} (in dashed) rather than 
purity, which we define as
\begin{equation}
    I = 1 - P = 1 - \frac{N_{\text{QSO,TRUE \& QSO,PRED}}}{N_{\text{QSO,PRED}}}.
\end{equation}

As validated in the plot, \texttt{DESI AL 2} has a significant improvement in completeness
at approximately the same (although marginally better) purity at each epoch compared to \texttt{DESI AL 1}.
\texttt{DESI AL 1} has a more impressive gain in completeness over the original \texttt{DESI VI}, although
it does suffer from decreased purity. Compared to \texttt{eBOSS DR12}, our final run, \texttt{DESI AL 2}, 
is comparable in completeness and purity, but uses less than a tenth as much data. 

The training curves shown in Figure~\ref{fig:purcom} show some oscillation and are not smooth. This
is because the training loss function used to train QuasarNET tries to minimize the distance
between the predicted line position and the true line position, equally weighting the
coarse and fine grained estimates (for more details see Equation~1 of \cite{busca_quasarnet_2018}). Classification uses only the coarse
grained confidence in line position, while redshift estimates use the both the coarse and 
fine grained output.
At some epochs the completeness or purity may fluctuate as confidence values change, but at the 
same epoch the redshift estimation may improve in accuracy. We only investigate the purity and completeness
here as we are only concerned with classification accuracy as opposed to redshift accuracy. 

\begin{figure}[tbp]
\centering 
\includegraphics[width=\textwidth]{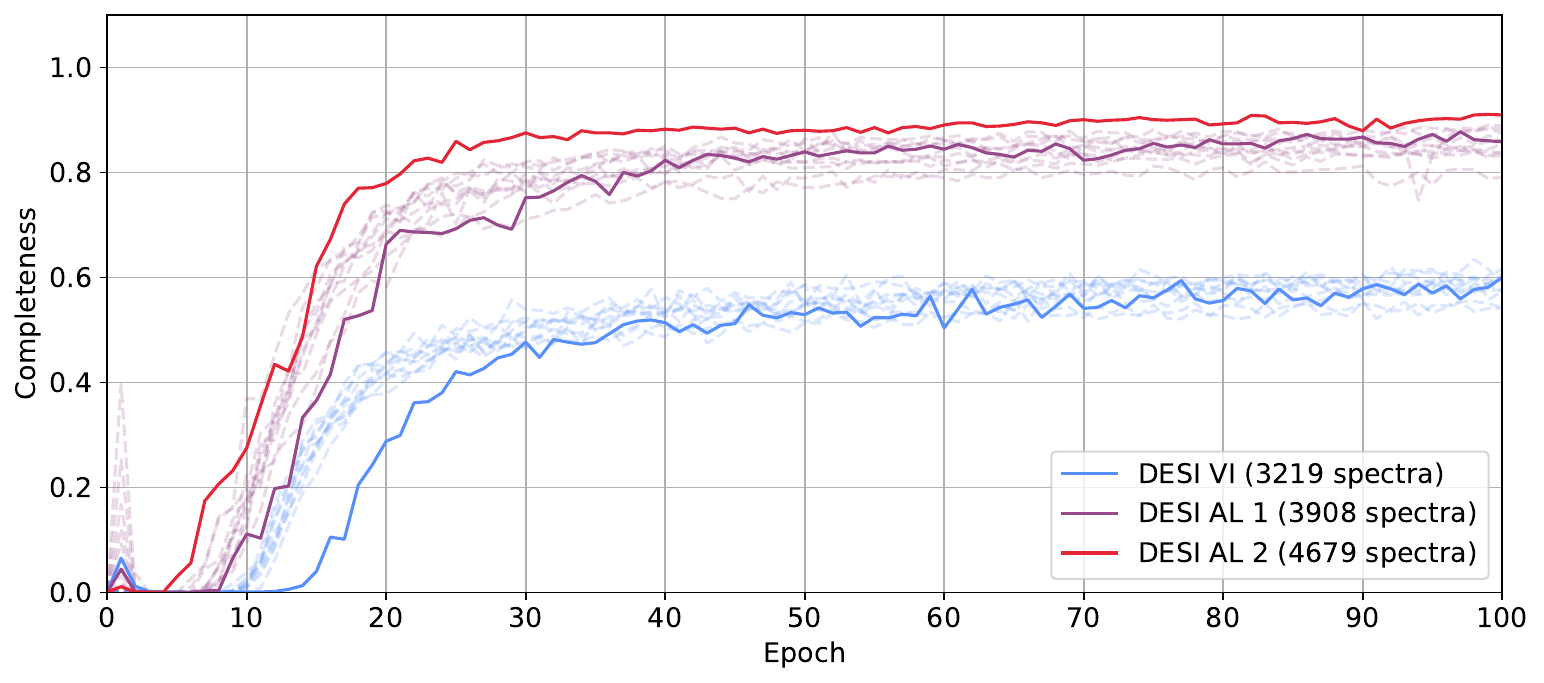}
\caption{\label{fig:bootstrap_loss} Completeness calculated on the validation dataset per epoch of training, with bootstraps. In solid we plot the completeness per epoch of training, the same data as in Figure~\ref{fig:purcom}. In dashed are bootstrapped realizations of the two smaller datasets. We bootstrap the smaller datasets to have the same amount of data as the largest (after the 2nd active learning run) in order to validate that the improvements when adding the active learning data at each step are, at least in part, due to the specific data chosen rather than the increase in dataset size. For both smaller datasets the bootstrapped datasets do not reach the same level of completeness, nor do they reach their maximum as quickly, as the final dataset. For clarity we have plotted only completeness, but purity follows the same trend.}
\end{figure}

It is possible that a larger dataset could explain some or all of the improvement at each active learning
step. In order to validate that our improvement is more significant than dataset size alone, we
bootstrap each of the two earlier runs, \texttt{DESI VI} and \texttt{DESI AL 1}, to have the same amount of data as the final assembled dataset. We perform 10 unique
bootstraps of each dataset, for 20 total unique datasets of the same size as the final dataset, and
then train for 100 epochs with the same parameters as the previous test. This bootstrapping simulates
naively adding additional spectra to each dataset drawn from the same distribution as that dataset. 

The results of this test are shown
in Figure~\ref{fig:bootstrap_loss}. For clarity we plot only the completeness, although purity
follows a similar trend. The solid lines represent the same completeness curves as plotted in Figure~\ref{fig:purcom}.
In dashed blue and dashed magenta are the 10 bootstrap runs for the \texttt{DESI VI} and \texttt{DESI AL 1} datasets respectively. Although the increased amount of data does have some effect on 
the training
it does not significantly alter the final best completeness values. This is 
not unexpected but it does validate that the active learning methodology has specifically selected
spectra that expand the quasar space covered by the training dataset, rather than simply sampling
similar quasars to those already covered by the training data. 

\begin{figure}[tbp]
\centering 
\includegraphics[width=\textwidth]{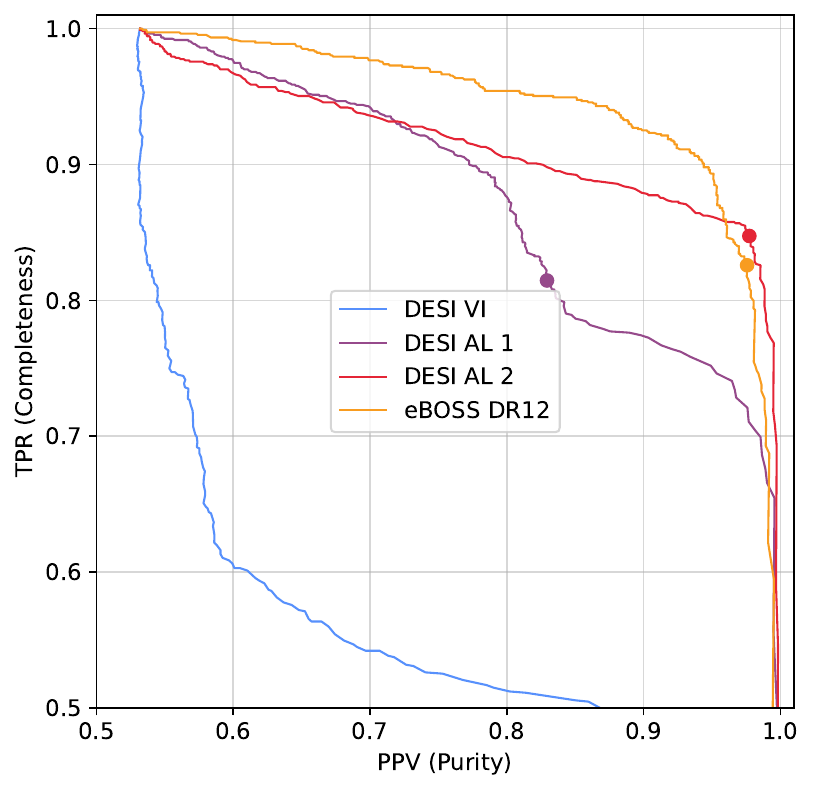}
\caption{\label{fig:roc} Plot of completeness versus purity for a variety of confidence thresholds ranging from 0 to 1 using the final produced weights file for each of the four datasets. At each confidence threshold we compute quasar classifications and the corresponding
completeness and purity on the validation dataset given those classifications. Colors match those of Fig.~\ref{fig:purcom}. A large dot on each curve (below the bottom of the plot for the \texttt{DESI VI} dataset) denotes the values at the standard 0.95 confidence threshold. At a confidence
threshold of 0.95, \texttt{DESI VI 2} outperforms the eBOSS weights file in
completeness with a similar purity, although there are some threshold values at which the opposite
is true.}
\end{figure}

For our active learning pipeline we used a confidence threshold of 0.95, meaning that any spectrum with 
at least one line with a confidence more than 0.95 was considered a quasar. This motivated the
choice used in the loss curves in Figure~\ref{fig:purcom} and Figure~\ref{fig:bootstrap_loss}, where the purity
and completeness per epoch were calculated using this same confidence threshold. 

It is also informative
to explore how the choice of confidence threshold changes the final results of the network. 
In Figure~\ref{fig:roc} we plot the purity and completeness against each other for the final networks
trained on each of the four datasets. The curves are computed by varying the confidence threshold used
for classification between 0 and 1, showing the trade off between improved
completeness and lower purity at lower thresholds and the opposite at higher thresholds. For example,
with a very low confidence threshold ($\sim 0.1$), the sample will be very complete but relatively
impure as almost everything will be classified a quasar.
A high threshold ($\sim 0.99$)
will generate a very pure but incomplete sample as only high confidence spectra are labeled quasars. 
Increasing the confidence threshold has the effect of moving along a curve from the top left of the figure
(low threshold) to the bottom right (high threshold). A perfect classifier is at the top right corner
of the figure (100\% pure and 100\% complete). 

Large dots in the figure
indicate the purity and completeness of each network at DESI's nominal confidence threshold of 0.95. 
At the nominal
threshold of 0.95 we have succeeded in improving the completeness of the final weights file, with almost
identical purity. Notably this effect does not hold true at all confidence thresholds. At lower thresholds
the \texttt{eBOSS DR12} weights file exceeds our newly trained weights files in completeness with similar or better
purity. This is due to our use of this confidence threshold when determining classification for
active learning, as we only pick spectra that are confused between classes when the constituent
networks of the ensemble are very confident in their own classifications. It is likely that
if we had used a different threshold we would see our network outperforming the eBOSS one at that
specific threshold, as we do with a threshold of 0.95. 

\subsection{Results on Unlabeled Data}
In this section we produce a weights file using the entirety of our DESI labeled data set as
a possible replacement for the current eBOSS trained weights. 
When we do this we do not split the assembled truth table, and use the
entire 6698 spectra in the truth table as input training data for the weights file. 
For validation purposes we use the finalized weights file and use it to classify and estimate 
redshifts for the entirety of the DESI DR1 ``Guadalupe'' release, a dataset of approximately 3.6 million 
unique spectra. In this section we will refer to the old weights file as the ``eBOSS weights'' and the
weights produced by this work as the ``DESI weights''.

\subsubsection{Classification and Consistency Results}

In DESI Survey Validation we reobserved objects an average of 3 times.  
We can use these repeat observations to test the consistency
of QuasarNET's output from exposure to exposure. Each exposure will have its own noise and sky
subtraction realizations, confounding the true signal in a unique way. 
Since the repeat exposures are of the same
object, we quantify improvements as those by which the network more consistently classifies the 
object as the same class, whether that be a quasar or not, in exposures of the same
object. Additionally for those objects which QuasarNET does consistently
classify as quasars we can compare the estimated redshift of each exposure to determine how
secure the redshift estimation is from exposure to exposure. 
\begin{figure}[tbp]
\centering 
\includegraphics[width=\textwidth]{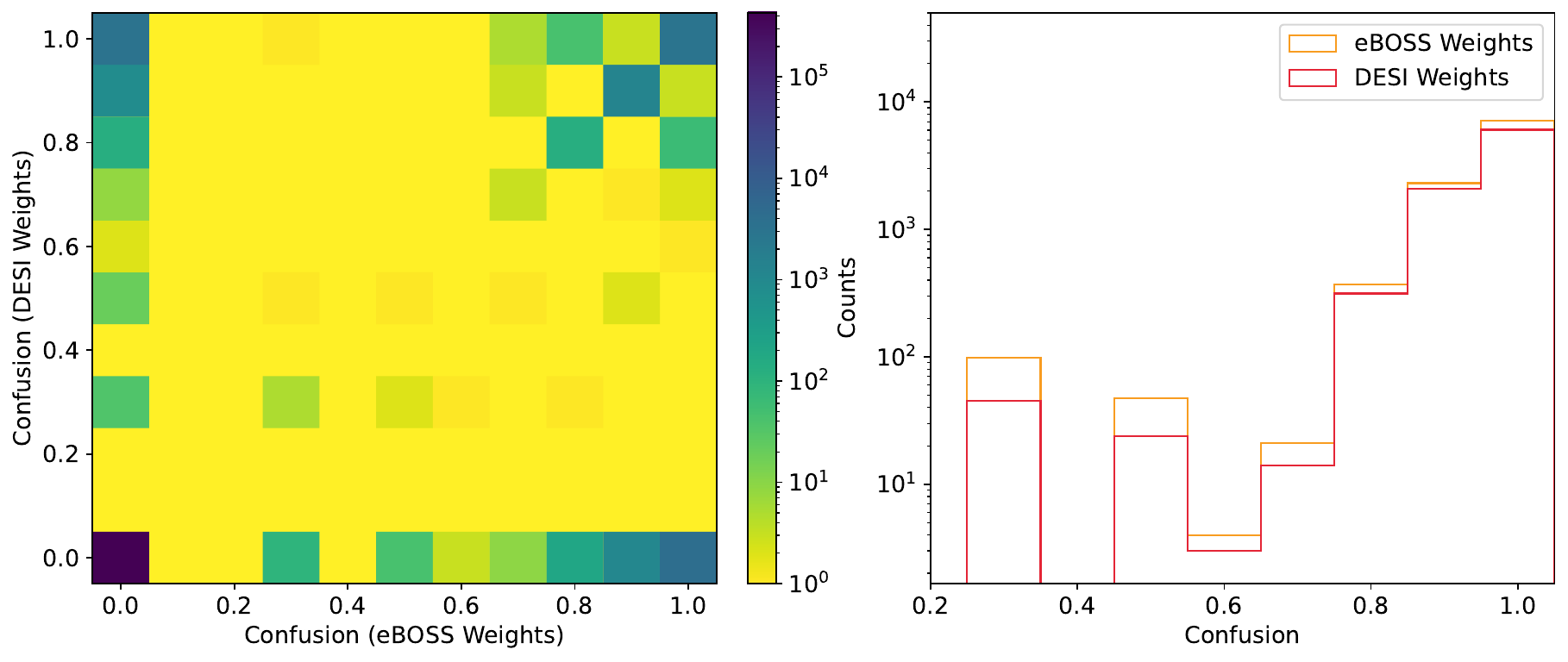}
\caption{\label{fig:confusion} Confusion of repeat exposures of the same object from the eBOSS and 
DESI weights. See text for details on how we compute confusion. In the left panel is a two dimensional histogram, plotting the eBOSS confusion values
vs the DESI confusion values. The bulk of the objects have no confusion, and thus a consistent
classification across exposures. The right panel is a one dimensional histogram of confusion values,
with the eBOSS results in blue and the DESI weights in orange. There is a notable decrease in 
all non-zero confusions values, with a corresponding increase in objects with
zero confusion. For clarity we do not plot the zero confusion objects in the right histogram.}
\end{figure}

We will use the entropy as defined in Eq. \ref{eq:entropy} with two classes (QSO and NOT QSO)
to determine classification confusion. The confusion is symmetric, for example when 60\% of the exposures 
classified as a quasar and 40\% are not the confusion value is the same as when 40\% are labeled quasars
and 60\% are not. The maximum confusion value is $-\log_2{0.5} = 1$ when the classifications are 
equally split between being a quasar and not. 
A confusion of 0 is only achieved when QuasarNET produces the same classification on all exposures.
In Figure~\ref{fig:confusion} we plot the estimated confusion values for each weights file. 
The left panel is a two dimensional plot of confusion values, with the vertical axis being the 
confusion from the new DESI weights (trained on \texttt{DESI AL 2}) and the horizontal axis being the confusion from
the eBOSS weights (used previously in DESI DR1, trained on \texttt{eBOSS DR12}), demonstrating where confusion values lie in this two dimensional space. 
The right panel overplots 1-D histograms of the two weights-files' results, underscoring the decrease in 
confusion values across all non-zero values with the new DESI weights. There is a corresponding slight
increase in 0 values in the new DESI weights, but the overwhelming number of repeat exposures in that
bin obfuscate this somewhat (note the logarithmic scaling on the vertical axis). 

\begin{figure}[tbp]
\centering 
\includegraphics[width=\textwidth]{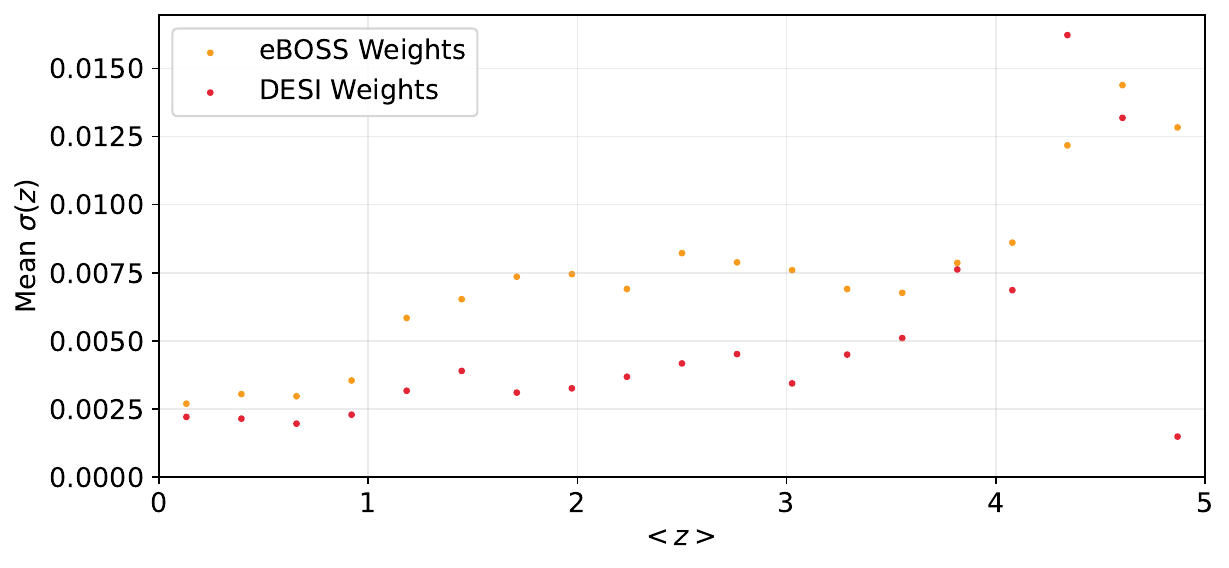}
\caption{\label{fig:std_z} The mean binned standard deviation in redshift estimate. For each object
we compute the standard deviation in redshift estimates from multiple exposures of the same object. Then we bin the objects in redshift, plotting the mean redshift of each bin and the mean of the standard deviations of the objects in that bin. In orange are the results from the prior eBOSS weights,
and in red are the results of the DESI weights. All bins above
$z = 4$ have low numbers of objects and corresponding higher variability. In all bins except one we produce
more consistent redshift estimates. }
\end{figure}

In Figure~\ref{fig:std_z} we show the results of the redshift consistency test, plotting
the standard deviation of redshift estimates of the same object versus the redshift. For clarity
we create bins in redshift, represented in the plot by their mean redshift value, and take the 
mean of the standard deviation values in that bin. In orange are the eBOSS weight's redshift estimates,
while below and in red are the DESI weight's estimates. The DESI weights produce consistently
lower variance in redshift estimates, representing an improvement in \emph{consistency} of redshift
estimates.

We would like to underscore that these results only validate that QuasarNET shows more consistency and 
less variability in results. This does not necessarily mean that QuasarNET is more accurate with the
new weights file, just that it gets the same result more often. Indeed, when comparing the redshift
estimates between the new and old weights files we discovered that both weights files share an issue
with redshift estimation.

\subsubsection{Redshift Results and Identifying an Issue with QuasarNET}
\begin{figure}[tbp]
\centering 
\includegraphics[width=0.8\textwidth]{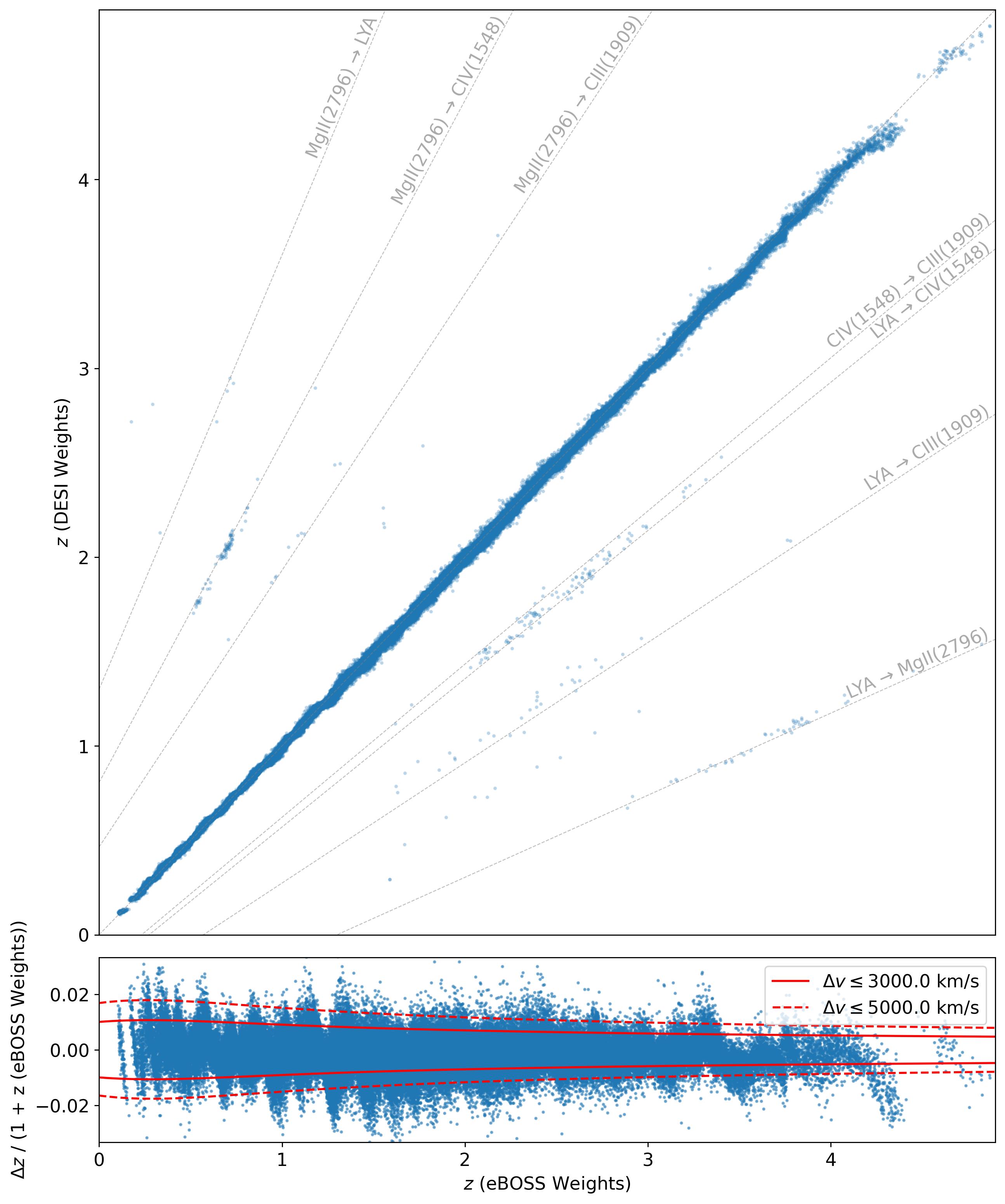}
\caption{\label{fig:redshift_redshift} Comparison between redshift estimates from the prior (eBOSS) weights file and the new (DESI) weights file. The top panel is a square plot of the two redshift estimates, most redshift estimates lie close to or along the diagonal where $z_{eBOSS} \approx z_{DESI}$. Some redshift estimates lie off the diagonal, and can be linked to ``line confusions'' where both networks are using the same observed-frame emission line to estimate redshift but differ in which rest-frame emission line they use for redshift estimation. Some of these line confusions are plotted as diagonal dashed grey lines, with the corresponding label at the right most edge of the line where the first line is the rest-frame line used by the eBOSS weights file and the second line is the rest-frame line used by the DESI weights file. In the bottom panel we plot $\Delta z / (1 + z_{eBOSS})$ to compare how significant the deviations from a perfect diagonal agreement are. The two cutoffs used in the text are plotted in dashed red ($\Delta v \leq 3000$ km/s) and solid red ($\Delta v \leq 5000$ km/s), with 89\% and 97.6\% of estimates falling within the respective curves. ``Oscillations'' around the diagonal are present in both panels, see text for an explanation.}
\end{figure}

Using the aforementioned DESI DR1 supplement we computed redshift estimates for all spectra
and compared them to the original classification and redshift estimates from the previous
weights file. This comparison is shown in
Figure~\ref{fig:redshift_redshift}. 
Plotted is a comparison of different redshift estimates between the eBOSS and DESI weights files. Only spectra
for which both weights files classify the object as a quasar are included in the plot, as QuasarNET
does not produce redshift estimates for non-quasar objects. The total number of spectra
included in the plot is 273,319, accounting for about 7.52\% of the total dataset. 

\begin{table}[tbp]
\centering
\caption{Table showing the number of spectra that have a velocity difference of $\leq 10000 $ km/s away from the labeled confusions on Figure~\ref{fig:redshift_redshift}.}
\begin{tabular}{|l|c|c|}
\hline
eBOSS Line Prediction & DESI Line Prediction & Number of Spectra \\ \hline
MgII(2796) & CIV(1548) & 76 \\
MgII(2796) & LYA & 5 \\
CIV(1548) & CIII(1909) & 70 \\
LYA & CIV(1548) & 89 \\
LYA & CIII(1909) & 17 \\
LYA & MgII(2796) & 42 \\
MgII(2796) & CIII(1909) & 8 \\ \hline
\end{tabular}
\label{table:confusions}
\end{table}
The bulk of the redshift estimates lie close to the main diagonal. In order to quantify how many
redshift estimates lie close to the diagonal and their level of agreement between the two
weights files we compute the recession velocity of the quasar from both redshift estimates and compare.
For recession velocity we use $v(z) = \frac{D_c(z)}{1 + z} H(z)$, where $D_c(z)$ is the
comoving distance, using the fiducial cosmology favored in \cite{2024arXiv240403002D}, 
which includes mild curvature, although the recession velocities are relatively robust
to changes in cosmological parameters. Using these recession velocities,  89\% of redshift 
estimates agree between the eBOSS and DESI weights within $\Delta v \leq 3000$km/s. 
97.6\% of the estimates agree within $\Delta v \leq 5000$km/s, indicating that an overwhelming number
of redshift estimates agree within reason between the weights files. 
This is shown in the bottom panel of Figure~\ref{fig:redshift_redshift}, where we plot 
$(z_{DESI} - z_{eBOSS}) / (1 + z_{eBOSS}) = \Delta z / (1 + z_{eBOSS})$ along with two cutoff lines
in dashed and solid red corresponding to the two aforementioned agreements respectively.
The redshift estimation of the eBOSS weights file for DESI was validated previously in 
\cite{farr_optimal_2020} and the significant agreement in redshift estimates reinforces that our new
DESI weights file is performing at a similar or better level of success to the previous weights file.

Some off-diagonal redshift changes can be explained by ``line-confusion,'' which is when two different 
weights files estimate redshift based on the position of the same emission line, but 
predict that the observer-frame emission
line corresponds to two different emission lines in the rest-frame. This swap in line classification
leads to an easily predictable change in redshift estimation. Some clearly identifiable line confusions
are plotted along diagonals in light, dashed, gray along with associated labels. The first line name in each 
label is the predicted associated rest-frame line from the eBOSS weights, where the second is the
prediction from the DESI weights. A summary of how many spectra whose velocity difference is 
$\leq 10000 $ away from the labeled confusion is shown in Table \ref{table:confusions}.
Many of these line confusions are a result of QuasarNET using only
a single emission line for redshift identification, and might be resolved if the spectrum shows
a second emission line to resolve the confusion. In some cases QuasarNET recognizes a second emission
line but has no way to use this information in its internal line identification. Note that the 
number of spectra that cluster around these confusions is comparatively small ($O(100)$) compared to
the total number of spectra used to generate the plot ($O(3000000)$).

In spectra that lie along the diagonal and have some measure of agreement in redshift estimates, 
we identify oscillations around perfect agreement. This can 
be seen prominently in the bottom panel where we plot the delta of the two redshift estimates. These
oscillations correspond exactly with the edges of the QuasarNET output boxes used for line position
estimation. Our active learning dataset, while correctly improving classification accuracy,
had the unintended side effect of revealing a deficiency in QuasarNET's redshift estimation. 
QuasarNET divides the output wavelength grid into 13 discrete boxes and has a tendency to clump
line position estimates towards the center of the boxes. This effect can be obfuscated by including more
data at all redshifts, correspondingly including more variety in line position and shape, which is
why the effect is less prominent in the eBOSS weights compared to the DESI weights, even though
it is still present. 
While this work did make this issue more prevalent, allowing us to discover and understand it,
it is undesirable to keep in the final weights file for DESI processing. For DESI's Year 3
processing we returned these oscillations to at or below the eBOSS level by including a larger
set of training data comprised of DESI spectra with VI labels cross matched from eBOSS. 
This improved QuasarNET's coverage at all redshifts, and mitigated the oscillations, but does
not solve the underlying problem which is inherent in the construction of QuasarNET. 

\section{Conclusions}
In this paper we have made significant changes to QuasarNET to improve its 
classification performance
for use in the spectral processing pipeline for DESI. Since QuasarNET was originally designed 
and trained on eBOSS data, we chose to make modifications to QuasarNET that would improve
its performance on DESI data. One of these changes included training on only 
labeled DESI data, a significantly reduced dataset compared to that of the previous
eBOSS trained weights file.

In order to accommodate the reduced amount of training data we modified the structure of QuasarNET by
including dropout layers and two additional convolution blocks. 
Both of these changes improve the generalizability of QuasarNET when trained on a reduced
training set, an inherent problem when using DESI data only for training. Since visual inspection
of DESI data has only been carried out on small subsamples of the full dataset, we only have
approximately 4600 spectra available to use for training, a significant decrease compared to
the eBOSS weights file. 

After modifying QuasarNET directly we then increased the amount of training data available by 
doing two additional rounds of visual inspection. In order to optimally distribute our limited
human effort and time we use an algorithm called active learning to select which data
to visually inspect. In our active learning pipeline we use bootstrapping to generate 200 separate
and unique networks from the same (minimal) training dataset. These networks classify each spectrum
into three categories, LOWZ, HIGHZ and NOT QSO. We then applied a novel outlier rejection step
to further optimize over time and effort to ensure that we only label spectra that are representative
of the bulk sample of quasars observed in the DESI survey. Our novel outlier
rejection step removes approximately 10\% of possible data for labeling. Ultimately we produced
a truth table of 6700 spectra for training, two spectra of which we removed due to 
data processing error, and used these spectra to train a new candidate weights file for the DESI
survey.

We demonstrate that the active learning methodology shows notable self improvement, improving the 
networks trained on the successive datasets produced by the algorithm. We also demonstrate that 
the active learning algorithm is selecting new spectra for training that expand the space of objects
covered by the training dataset. We use the final product weights file, trained after two active learning
iterations, as a candidate new DESI weights file. 

When used on DESI data with DESI classification parameters
the new weights file has somewhat better completeness with a similar purity,
reducing the number of false negative quasar identifications. When comparing different
exposures of the same object our new weights file produces more consistent classifications, more
often classifying objects the same way in different exposures of the same object. For objects
that are consistently classified as a quasar our new weights file also has a significantly lower
spread in redshift estimates. 

However, for objects classified as a quasar in both the old and new weights
files, we note that even though the majority of our redshift estimates agree with the eBOSS weights
predictions our weights file revealed minor issues with QuasarNET's redshift estimation. QuasarNET
tends to clump its line position estimates, in turn leading to wiggles or oscillations in redshift
estimates. In this paper we focused exclusively on selecting spectra that would improve QuasarNET's
classification accuracy, but as a result discovered that our resulting training sample does not
cover the entire redshift range. Further work into both understanding and removing 
these redshift wiggles
is currently underway. For DESI's Year 3 processing
we have used the understanding of this problem, gained in this paper, to mitigate these issues by
including an additional set of DESI training data cross matched from eBOSS, thus returning the oscillations to the eBOSS levels while maintaining the improved classification accuracy gained from the Active Learning
pipeline.


\acknowledgments
The work of Dylan Green and David Kirkby was supported by the U.S. Department of Energy,
Office of Science, Office of High Energy Physics, under Award No. DE-SC0009920.

This material is based upon work supported by the U.S. Department of Energy (DOE), Office of Science, Office of High-Energy Physics, under Contract No. DE–AC02–05CH11231, and by the National Energy Research Scientific Computing Center, a DOE Office of Science User Facility under the same contract. Additional support for DESI was provided by the U.S. National Science Foundation (NSF), Division of Astronomical Sciences under Contract No. AST-0950945 to the NSF’s National Optical-Infrared Astronomy Research Laboratory; the Science and Technology Facilities Council of the United Kingdom; the Gordon and Betty Moore Foundation; the Heising-Simons Foundation; the French Alternative Energies and Atomic Energy Commission (CEA); the National Council of Humanities, Science and Technology of Mexico (CONAHCYT); the Ministry of Science, Innovation and Universities of Spain (MICIU/AEI/10.13039/501100011033), and by the DESI Member Institutions: \url{https://www.desi.lbl.gov/collaborating-institutions}. Any opinions, findings, and conclusions or recommendations expressed in this material are those of the author(s) and do not necessarily reflect the views of the U. S. National Science Foundation, the U. S. Department of Energy, or any of the listed funding agencies.

The authors are honored to be permitted to conduct scientific research on Iolkam Du’ag (Kitt Peak), a mountain with particular significance to the Tohono O’odham Nation.

\noindent \textbf{Data Availability} All data and python code used to generate the 
plots in  this paper are accessible at \url{https://doi.org/10.5281/zenodo.15328537}.


\bibliographystyle{JHEP.bst}
\bibliography{refs.bib}

\end{document}

%% file: qnet_authors.tex
\author[1]{{Dylan~Green}\orcidlink{0000-0002-0676-3661},}
\author[1]{{David~Kirkby}\orcidlink{0000-0002-8828-5463},}
\author[2]{{J.~Aguilar},}
\author[3]{{S.~Ahlen}\orcidlink{0000-0001-6098-7247},}
\author[4,5]{{D.~M.~Alexander}\orcidlink{0000-0002-5896-6313},}
\author[6]{{E.~Armengaud}\orcidlink{0000-0001-7600-5148},}
\author[2]{{S.~Bailey}\orcidlink{0000-0003-4162-6619},}
\author[2]{{A.~Bault}\orcidlink{0000-0002-9964-1005},}
\author[7,8]{{D.~Bianchi}\orcidlink{0000-0001-9712-0006},}
\author[2]{{A.~Brodzeller}\orcidlink{0000-0002-8934-0954},}
\author[9]{{D.~Brooks},}
\author[2]{{T.~Claybaugh},}
\author[2]{{R.~de Belsunce}\orcidlink{0000-0003-3660-4028},}
\author[10]{{A.~de la Macorra}\orcidlink{0000-0002-1769-1640},}
\author[9]{{P.~Doel},}
\author[11]{{V.~A.~Fawcett}\orcidlink{0000-0003-1251-532X},}
\author[2,12]{{S.~Ferraro}\orcidlink{0000-0003-4992-7854},}
\author[13]{{A.~Font-Ribera}\orcidlink{0000-0002-3033-7312},}
\author[14,15]{{J.~E.~Forero-Romero}\orcidlink{0000-0002-2890-3725},}
\author[16,17,18]{{E.~Gaztañaga},}
\author[2]{{S.~Gontcho A Gontcho}\orcidlink{0000-0003-3142-233X},}
\author[19]{{G.~Gutierrez},}
\author[20]{{M.~Ishak}\orcidlink{0000-0002-6024-466X},}
\author[21]{{S.~Juneau}\orcidlink{0000-0002-0000-2394},}
\author[22]{{R.~Kehoe},}
\author[2]{{T.~Kisner}\orcidlink{0000-0003-3510-7134},}
\author[2]{{A.~Kremin}\orcidlink{0000-0001-6356-7424},}
\author[2]{{A.~Lambert},}
\author[2]{{M.~Landriau}\orcidlink{0000-0003-1838-8528},}
\author[23]{{L.~Le~Guillou}\orcidlink{0000-0001-7178-8868},}
\author[2]{{M.~E.~Levi}\orcidlink{0000-0003-1887-1018},}
\author[24,13]{{M.~Manera}\orcidlink{0000-0003-4962-8934},}
\author[21]{{A.~Meisner}\orcidlink{0000-0002-1125-7384},}
\author[25,13]{{R.~Miquel},}
\author[26]{{J.~Moustakas}\orcidlink{0000-0002-2733-4559},}
\author[27]{{A.~D.~Myers},}
\author[6,2]{{N.~Palanque-Delabrouille}\orcidlink{0000-0003-3188-784X},}
\author[28]{{F.~Prada}\orcidlink{0000-0001-7145-8674},}
\author[29]{{I.~P\'erez-R\`afols}\orcidlink{0000-0001-6979-0125},}
\author[30]{{G.~Rossi},}
\author[31]{{E.~Sanchez}\orcidlink{0000-0002-9646-8198},}
\author[32]{{C.~Saulder}\orcidlink{0000-0002-0408-5633},}
\author[2]{{D.~Schlegel},}
\author[33,34]{{M.~Schubnell},}
\author[35]{{H.~Seo}\orcidlink{0000-0002-6588-3508},}
\author[36,37]{{F.~Sinigaglia}\orcidlink{0000-0002-0639-8043},}
\author[21]{{D.~Sprayberry},}
\author[6]{{T.~Tan}\orcidlink{0000-0001-8289-1481},}
\author[34]{{G.~Tarl\'{e}}\orcidlink{0000-0003-1704-0781},}
\author[21]{{B.~A.~Weaver},}
\author[17]{{S.~Youles}\orcidlink{0000-0002-7520-5911},}
\author[38]{{J.~Yu}\orcidlink{0009-0001-7217-8006},}
\author[2]{{R.~Zhou}\orcidlink{0000-0001-5381-4372},}
\author[39]{{H.~Zou}\orcidlink{0000-0002-6684-3997},}

\affiliation[1]{Department of Physics and Astronomy, University of California, Irvine, 92697, USA}
\affiliation[2]{Lawrence Berkeley National Laboratory, 1 Cyclotron Road, Berkeley, CA 94720, USA}
\affiliation[3]{Department of Physics, Boston University, 590 Commonwealth Avenue, Boston, MA 02215 USA}
\affiliation[4]{Centre for Extragalactic Astronomy, Department of Physics, Durham University, South Road, Durham, DH1 3LE, UK}
\affiliation[5]{Institute for Computational Cosmology, Department of Physics, Durham University, South Road, Durham DH1 3LE, UK}
\affiliation[6]{IRFU, CEA, Universit\'{e} Paris-Saclay, F-91191 Gif-sur-Yvette, France}
\affiliation[7]{Dipartimento di Fisica ``Aldo Pontremoli'', Universit\`a degli Studi di Milano, Via Celoria 16, I-20133 Milano, Italy}
\affiliation[8]{INAF-Osservatorio Astronomico di Brera, Via Brera 28, 20122 Milano, Italy}
\affiliation[9]{Department of Physics \& Astronomy, University College London, Gower Street, London, WC1E 6BT, UK}
\affiliation[10]{Instituto de F\'{\i}sica, Universidad Nacional Aut\'{o}noma de M\'{e}xico,  Circuito de la Investigaci\'{o}n Cient\'{\i}fica, Ciudad Universitaria, Cd. de M\'{e}xico  C.~P.~04510,  M\'{e}xico}
\affiliation[11]{School of Mathematics, Statistics and Physics, Newcastle University, Newcastle upon Tyne, NE1 7RU, UK}
\affiliation[12]{University of California, Berkeley, 110 Sproul Hall \#5800 Berkeley, CA 94720, USA}
\affiliation[13]{Institut de F\'{i}sica d’Altes Energies (IFAE), The Barcelona Institute of Science and Technology, Edifici Cn, Campus UAB, 08193, Bellaterra (Barcelona), Spain}
\affiliation[14]{Departamento de F\'isica, Universidad de los Andes, Cra. 1 No. 18A-10, Edificio Ip, CP 111711, Bogot\'a, Colombia}
\affiliation[15]{Observatorio Astron\'omico, Universidad de los Andes, Cra. 1 No. 18A-10, Edificio H, CP 111711 Bogot\'a, Colombia}
\affiliation[16]{Institut d'Estudis Espacials de Catalunya (IEEC), c/ Esteve Terradas 1, Edifici RDIT, Campus PMT-UPC, 08860 Castelldefels, Spain}
\affiliation[17]{Institute of Cosmology and Gravitation, University of Portsmouth, Dennis Sciama Building, Portsmouth, PO1 3FX, UK}
\affiliation[18]{Institute of Space Sciences, ICE-CSIC, Campus UAB, Carrer de Can Magrans s/n, 08913 Bellaterra, Barcelona, Spain}
\affiliation[19]{Fermi National Accelerator Laboratory, PO Box 500, Batavia, IL 60510, USA}
\affiliation[20]{Department of Physics, The University of Texas at Dallas, 800 W. Campbell Rd., Richardson, TX 75080, USA}
\affiliation[21]{NSF NOIRLab, 950 N. Cherry Ave., Tucson, AZ 85719, USA}
\affiliation[22]{Department of Physics, Southern Methodist University, 3215 Daniel Avenue, Dallas, TX 75275, USA}
\affiliation[23]{Sorbonne Universit\'{e}, CNRS/IN2P3, Laboratoire de Physique Nucl\'{e}aire et de Hautes Energies (LPNHE), FR-75005 Paris, France}
\affiliation[24]{Departament de F\'{i}sica, Serra H\'{u}nter, Universitat Aut\`{o}noma de Barcelona, 08193 Bellaterra (Barcelona), Spain}
\affiliation[25]{Instituci\'{o} Catalana de Recerca i Estudis Avan\c{c}ats, Passeig de Llu\'{\i}s Companys, 23, 08010 Barcelona, Spain}
\affiliation[26]{Department of Physics and Astronomy, Siena College, 515 Loudon Road, Loudonville, NY 12211, USA}
\affiliation[27]{Department of Physics \& Astronomy, University  of Wyoming, 1000 E. University, Dept.~3905, Laramie, WY 82071, USA}
\affiliation[28]{Instituto de Astrof\'{i}sica de Andaluc\'{i}a (CSIC), Glorieta de la Astronom\'{i}a, s/n, E-18008 Granada, Spain}
\affiliation[29]{Departament de F\'isica, EEBE, Universitat Polit\`ecnica de Catalunya, c/Eduard Maristany 10, 08930 Barcelona, Spain}
\affiliation[30]{Department of Physics and Astronomy, Sejong University, 209 Neungdong-ro, Gwangjin-gu, Seoul 05006, Republic of Korea}
\affiliation[31]{CIEMAT, Avenida Complutense 40, E-28040 Madrid, Spain}
\affiliation[32]{Max Planck Institute for Extraterrestrial Physics, Gie\ss enbachstra\ss e 1, 85748 Garching, Germany}
\affiliation[33]{Department of Physics, University of Michigan, 450 Church Street, Ann Arbor, MI 48109, USA}
\affiliation[34]{University of Michigan, 500 S. State Street, Ann Arbor, MI 48109, USA}
\affiliation[35]{Department of Physics \& Astronomy, Ohio University, 139 University Terrace, Athens, OH 45701, USA}
\affiliation[36]{Departamento de Astrof\'{\i}sica, Universidad de La Laguna (ULL), E-38206, La Laguna, Tenerife, Spain}
\affiliation[37]{Instituto de Astrof\'{\i}sica de Canarias, C/ V\'{\i}a L\'{a}ctea, s/n, E-38205 La Laguna, Tenerife, Spain}
\affiliation[38]{Institute of Physics, Laboratory of Astrophysics, \'{E}cole Polytechnique F\'{e}d\'{e}rale de Lausanne (EPFL), Observatoire de Sauverny, Chemin Pegasi 51, CH-1290 Versoix, Switzerland}
\affiliation[39]{National Astronomical Observatories, Chinese Academy of Sciences, A20 Datun Road, Chaoyang District, Beijing, 100101, P.~R.~China}